# FLOWER: Flow-Oriented Entity-Relationship Tool

Dmitry Moskalev[1][0009-0003-0076-0536]

[1] MTC Web Services, Novosibirsk, Russia
invite.dmitry@gmail.com

**Abstract.** Exploring relationships across data sources is a crucial optimization for entities recognition. Since databases can store big amount of information with synthetic and organic data, serving all quantity of objects correctly is an important task to deal with. However, the decision of how to construct entity relationship model is associated with human factor. In this paper, we present flow-oriented entity-relationship tool. This is first and unique end-to-end solution that eliminates routine and resource-intensive problems of processing, creating and visualizing both of explicit and implicit dependencies for prominent SQL dialects on-the-fly. Once launched, FLOWER automatically detects built-in constraints and starting to create own correct and necessary one using dynamic sampling and robust data analysis techniques. This approach applies to improve entity-relationship model and data storytelling to better understand the foundation of data and get unseen insights from DB sources using SQL or natural language. Evaluated on state-of-the-art STATS benchmark, experiments show that FLOWER is superior to reservoir sampling by 2.4× for distribution representation and 2.6× for constraint learning with 2.15× acceleration. For data storytelling, our tool archives 1.19× for accuracy enhance with 1.86× context decrease compare to LLM. Presented tool is also support 23 languages and compatible with both of CPU and GPU. Those results show that FLOWER can manage with real-world data a way better to ensure with quality, scalability and applicability for different use-cases.

## 1 Introduction

Entity-relationship model is a crucial conceptual framework to design manageable database logical components from real-world data abstracts. It is an important feature for databases that can operate independently of physical data model.

The main conceptual properties of ER-model includes definition, attributes and linking across all possible, but relevant entities from provided data sources. The common entities kinds of linking across DB are columns, tables and schemas. Creating ER-model is the popular way for translating business requirements to understand data in a more appropriate manner across relational DBMS. Building relationships across the entities plays a crucial role to define database structure correctly [1] and it is widely used in different domains [2, 3] for optimization purposes. Understanding of how to create entity-relationship model can bring many advantages for companies and their business structures to architecture, decompose complex domains into logical modules and validate the results.

For additional purposes or better understanding, ER-model can be visualized using entity relationship diagram (ERD). It is often used to formalize business rules and integrity constraints for different use-cases such as data dependencies and normalization requirements. Nowadays, companies try to use different ways to adapt and provide actual and relevant information in time. However, usually it requires additional processes and techniques. For those reasons, we propose tool that includes ER-model creation, construction and easily modification to use actual entities. It also improves interaction with databases for both of internal and external scenarios.

In sum, the key contributions of this paper include:
- Understand initial data distribution and how to process it in the dynamically manner to handle real-world data.
- Make semantics and columns data processing using NLP to get accurate results using own decision-making system.
- Evaluate constraint learning accuracy using state-of-the-art (SOTA) and own benchmarks.

## 2 Related Work

The foundational ER-model has generated extensive academic and industrial research since its creation. In recent time there is a meaningful interest to manage with relationships [4] and build the right ER-models [5] including constraint identification using statistical [6], machine learning [7] and analyzing database-backed [8] techniques for DB sources. While it can be enough for simple scenarios when there is only single schema and simple data distribution, it remains challenging of how not to miss potential dependencies in real database scenarios.

As far as market grows, many researches show that more and more real-world data consist of distributions with mixture [9], biased [10] or shift [11] patterns. However, sampling from initial data can led to distribution changes or can be slow in terms of complexity when we talk about big data processing.

Some articles also proposed to do a processing with semantic of entities [12] to improve the pipeline and get more insights from entities thought constraint learning. While it covers a new aspect, relying only on naming may lead to mistakes to define real dependencies due to subjective column labeling by authors or database administrators. Therefore, challenges about verification of diagram correctness remains the same.

Finally yet importantly is adoption to LLMs for Text2SQL (NL2SQL) tasks, when we need to generate correct SQL based on provided input on natural language. As it shows on recent research [13], providing only relevant information improves the number of correct query generations. However, it is still an open question of how to get neither more nor less entities from the schema to get correct SQL and prevent redundant information not to let LLM to hallucinate [14]. Thus, it is also an open question for possible integration into data platforms.

To address these challenges, we present FLOWER – tool that targets for real-world database use-cases. Our solution focuses on entities constraints recognition as pipeline of dynamic data analysis and natural language processing (NLP) techniques. FLOWER automatically processes information from databases and define implicit and explicit constraints based on own decision-making system from stored data. The importance of creating and optimizing dependencies in terms of ER-models is to help analysts and database administrators with table data annotation and self-checking. It also improves data storytelling and correct query generation to handle with OLTP and OLAP scenarios simultaneously.

## 3 FLOWER

This paper presents FLOWER – a flexible tool for learning, applying and visualizing implicit and explicit constraints to all modern SQL dialects that rely on ER-models architecture.

Given a database management system (DBMS) the tool aims to make all necessary constraints using decision-making pipeline automatization. Thus, it creates or improves current ER-model and provides more accurate and faster interaction with DB using natural language. The tool extends general pipeline without limitations of using only columns data or just entities semantics. Thus, it can archive better accuracy in constraint learning and balance between semantics and data since usually we cope with not empty entities.

In the following, it is proposed as a new approach to manage with described limitations. FLOWER mainly focuses on adaptability and scalability to create, replace or extend default ER-model within given DB source. In addition, current tool provides tunable hyperparameters to adapt for few scenarios of usage depending on customer needs. Users can choose balance or accuracy as a prior mode. The implicit dependencies also have two major steps: learning and inference. Finally, there is an option to visualize dependencies to understand if there are relevant and to export it in different user-friendly formats.

We will describe each step in more details as part of the following subsections of current research paper.

## 3.1 Architectural Overview

Figure 1 shows common end-to-end pipeline of FLOWER tool.

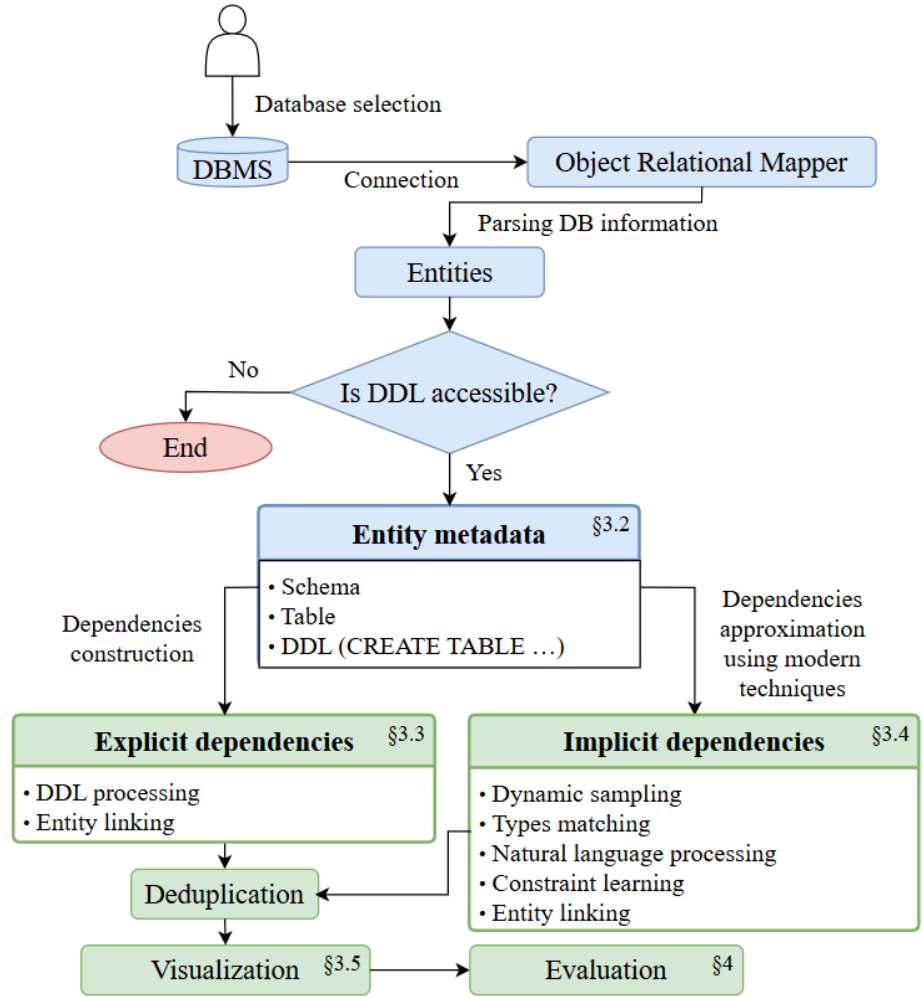

**Fig. 1.** Common FLOWER tool pipeline.

We start with connecting to one of DBMS dialects and analyze provided entities using object relational mapper (ORM). Secondly, for each collected entity we analyze data definition language (DDL) to catch explicit dependencies and move into sampling and NLP steps to find implicit dependencies. After that, we finish with entities processing with deduplication and removing implicit dependencies from common list if entity dependency have already selected as explicit one. In final step, we draw dependencies across all DB source and do evaluation if it is necessary.

## 3.2 Entities Preprocessing

In FLOWER, it is required to process with real data, so firstly we need to connect to existing DBMS using ORM or other connectors. The solution supports many popular dialects of relational databases.

After connection to DB source, we evaluate each table entity from DB to understand if it is accessible to get DDL or not. Depending on that, we collect full entity list of tables for further processing.

Notable, that based on data definition language structure we have two main cases. First one is explicit dependencies when in DDL we have well-written references. Another one is implicit dependencies for DDL without references when we need to approximate those connections between entities across the database.

## 3.3 Explicit Dependencies

Explicit dependencies strongly rely on entity structure that is why we need it from previous step. Giving with DDL, we try to identify labeled constraints based on dialect rules. Here is the example for PostgreSQL:

```
CREATE TABLE Orders(
    Id SERIAL PRIMARY KEY,
    CustomerId INTEGER,
    Quantity INTEGER,
    FOREIGN KEY (CustomerId) REFERENCES Customers (Id)
);
```

Here we have reference condition that FLOWER will catch and understand from-to entities, in this example it will be "Customers" as main table and "Id" as main column with explicit connection from "CustomerId" of "Orders" table. Notable, that in case of few constraints in one DDL tool will also manage with it. However, in common case companies do not have well-labeled DDL since it is not required by most of modern DMBS and we might have this example:

```
CREATE TABLE Orders(
    Id SERIAL PRIMARY KEY,
    CustomerId INTEGER,
    Quantity INTEGER
);
```

For those scenarios, we need to focus on implicit dependencies and try to reproduce missing or required constraints.

## 3.4 Implicit Dependencies

The two main steps of implicit dependencies are learning and inference. In learning stage, we dynamically select appropriate part of initial data using sampling to understand the distribution and common data pattern. Then we use NLP techniques to remove potential noise from column names and process with cleared text to match relationship between columns. The inference step is to evaluate the processed data using own adaptive metrics to understand is entities depending on each other or not.

**Dynamic Sampling.** The main purpose of dynamic sampling is to represent initial distribution in common case. It requires some flexibility, because we do not know the domain size and distribution pattern of columns. To manage with it we decided to pay attention not only for all count of rows, but also for unique amount of column data and coefficient of scalability. The minimum number of rows when tool starts to sample based on accuracy and latency tests was selected as 15000, so initially FLOWER requires 2× less data that reservoir sampling in PostgreSQL with default grand unified configuration (GUC) value using for planner improvements. Our tool uses fully random unbiased sampling with dynamic amount of rows (Eq. 1):

$$rows_{sampled} = \begin{cases} rows_{all}, if\ rows_{all} \leq rows_{min} \\ rows_{min} + \sqrt{\frac{rows_{uq}}{rows_{all}}} * rows_{min} + \frac{rows_{all}}{\sqrt{rows_{min}}}, otherwise \end{cases}, \quad (1)$$

where $rows_{min}$ – minimum number of rows for sampling, $rows_{uq}$ – number of column unique values, $rows_{all}$ – amount of all rows in column. If $rows_{sampled}$ after calculation is fractional number, then we use built-in function to convert it into integer value. Figure 2 presents the percentage of sampled and initial rows number.

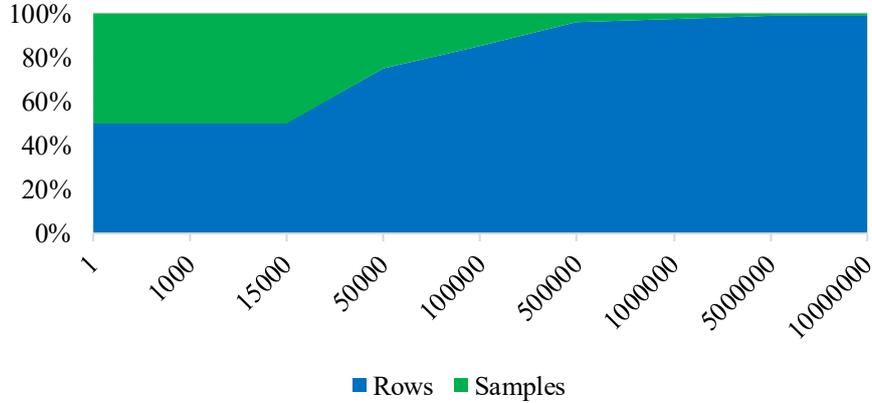

**Fig. 2.** Percentage of sampled and initial number of rows (normal distribution).

For testing purposes we use own dynamic sampling approach and standard reservoir sampling as baseline since it goes as built-in solution in some DBMS.

We used 10 launches per each test in Table 1 to present unbiased results of average sum of squared errors (SSE) metric. It is proposed to calculate difference between each observation and its group's mean in histogram bin to understand overall distribution similarity. Each launch is stand for rows amount calculation using dynamic sampling and next approximation of histogram bins error.

Figure 3 shows example for both of methods to understand distribution of sampled data.

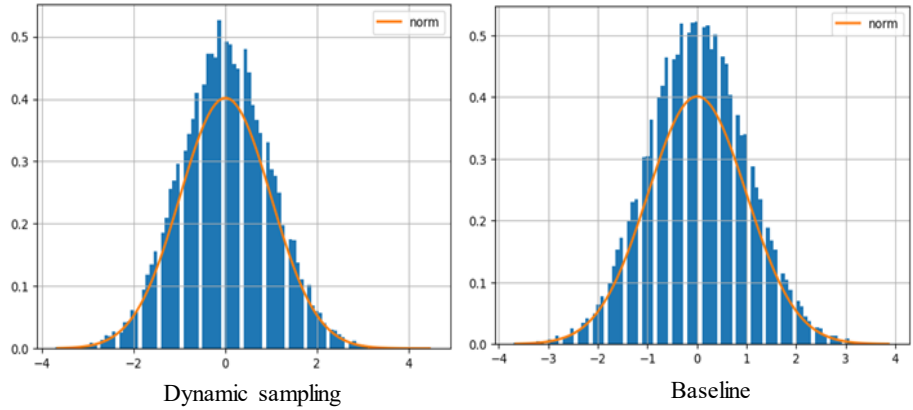

Dynamic sampling          Baseline

**Fig. 3.** Initial normal distribution match for *dynamic sampling* and *baseline* approaches.

Based on results current solution is better handle initial distribution for large number or rows in column.

**Table 1.** Dynamic sampling for normal distribution.

| Rows | Dynamic sampling | %, rows | Baseline | %, rows | $SSE_{adaptive}$ | $SSE_{baseline}$ |
|---|---|---|---|---|---|---|
| 100000 | *16253* | 16.25 | 30000 | 30 | 1.08 | 0.91 |
| 1000000 | *23311* | 2.33 | 30000 | 3 | 0.81 | 0.81 |
| 10000000 | 96697 | 0.97 | 30000 | 0.3 | *0.52* | 0.85 |
| 100000000 | 831512 | 0.83 | 30000 | 0.03 | *0.14* | 0.84 |

We obtained ~2.4× SSE metric improvement on average when divided summed results of $SSE_{baseline}$ and $SSE_{adaptive}$ columns.

Before moving to NLP it is necessary to check classes of pairs according to PostgreSQL example (Table 2).

**Table 2.** Classification for column types in PostgreSQL

| Class | Types |
|---|---|
| Digits | integer, decimal, numeric, real, double precision, serial, etc. |
| Money | money |
| Character | character varying, varchar, character, char, text |
| Binary | bytea |
| Data | timestamp, date, time, interval |
| Boolean | boolean, bit, bit varying |
| Geometric | line, point, lseg, box, path, polygon, circle |
| Network | cidr, inet, macaddr |

**Natural Language Processing.** After make the subset from initial data, FLOWER continues with text processing using NLP techniques and tokenization to split column names into tokens with current adaptation for 23 languages. After that, for each column we start to synonymize tokens using adaptive count based on tokens number in naming (Eq. 2):

$$synonyms_{amount} = \frac{confidence}{confidence_{coeff} * tokens_{number}}, \quad (2)$$

where $confidence$ is metric to understand is there implicit connection or not, $confidence_{coeff}$ is coefficient that adapt $confidence$ based on intersection between synonyms and $tokens_{number}$ is amount of tokens after tokenization. If $tokens_{number}$ is greater than one, then we split integer number of synonyms for each token. Figure 4 shows natural language processing example.

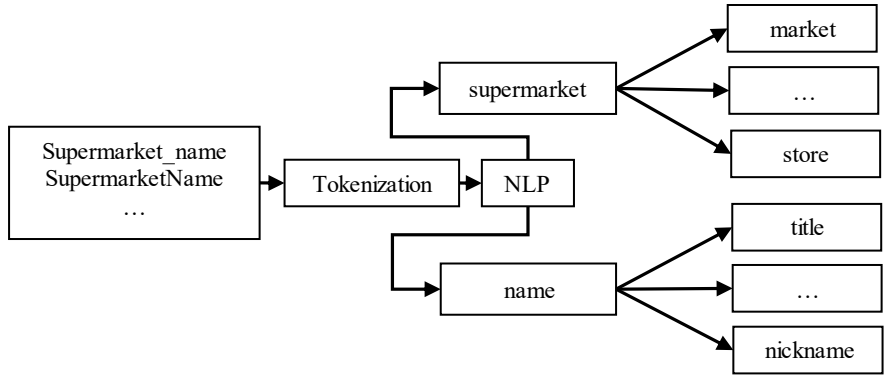

**Fig. 4.** Natural language processing for initial column name.

Then we start to adapt $confidence$ metric (Eq. 5) based of how many synonyms are between column pairs based on formula. Thus, we take into account not only intersection of elements from calculated amount of rows elements after dynamic sampling (Eq. 3), but also potential column names similarity (Eq. 4):

$$rows_{intersection} = \begin{cases} \frac{|rows_i \cap rows_j|}{|rows_j|}, if\ |rows_j| \neq 0 \\ 0, else \end{cases}, \quad (3)$$

$$synonyms_{\text{intersection}} = |synonyms_i \cap synonyms_j|, \quad (4)$$

$$confidence = confidence - (synonyms_{\text{intersection}} * confidence_{coeff}), \quad (5)$$

where $i$ and $j$ are column indexes in table and $i \neq j$.

After that, we can evaluate if there an implicit connection if it fits criteria:
- $rows_{intersection} \geq confidence$
- At least one of column pair is a primary key
- Types of compared columns are from the same class (only for pairs without common synonyms)

Finally, tool makes deduplication for cases when we already have these dependencies in explicit list and concatenate processed entities for next visualization.

### 3.5 Visualization

This section presents results of constraint learning and entities recognition using FLOWER in real-world DB source. Each entity uses pattern name of "schema.table". On Figure 5, the solid line represents explicit dependencies between entities. Dashed entity line means that table is empty.

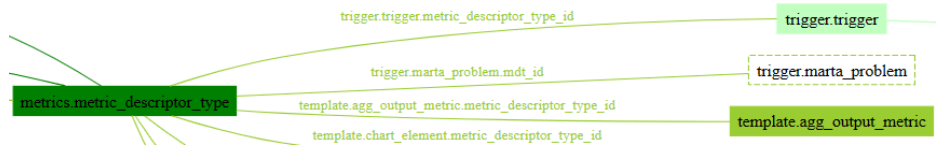

**Fig. 5.** Explicit connections visualization using FLOWER.

On Figure 6, we use dashed line for implicit constraints with corresponded metric to show user dependency confidence between entities linking.

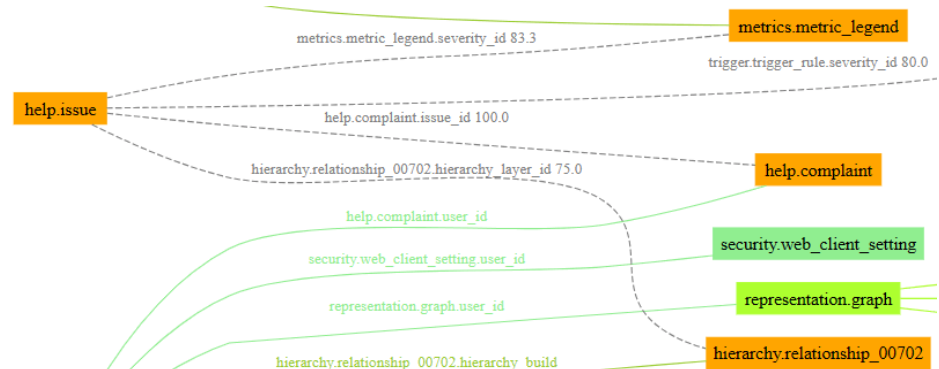

**Fig. 6.** Implicit connections visualization using FLOWER.

Visualization and metrics are useful to decide is it worth to use provided ER-model or change some output dependencies manually if it is necessary.

## 4 Evaluation

To evaluate FLOWER solution, we applied it to STATS benchmark based on real-data dataset from Stats Stack Exchange network [15] with built-in ER-model and 1029842 rows from 8 tables in total. We use our tool and outperforming modern LLM (DeepSeek R1 model with 32B parameters) that also support reasoning as baseline for testing purposes. As for hardware, we use Intel Core i7-13700H as CPU and NVIDIA A100 as GPU. The main goal of experiments is to show how better FLOWER is for constraint learning and data storytelling. We compared our solution for ERD construction usage in two scenarios. It show that tool is not hallucinate based on metadata or provided context and can handle different use-cases.

The accuracy metric (Eq. 6) we use to check correctness of ERD is:

$$accuracy = \frac{GT}{predicted}, \qquad (2)$$

where $GT$ – manually collected explicit dependencies from benchmark schema, $predicted$ – implicit dependencies collected with FLOWER tool.

Figure 7 shows the results of evaluations for correct dependencies approximation. We sum up all dependencies provided with our solution and LLM to understand is there any hallucinations when dependencies are not in ground truth.

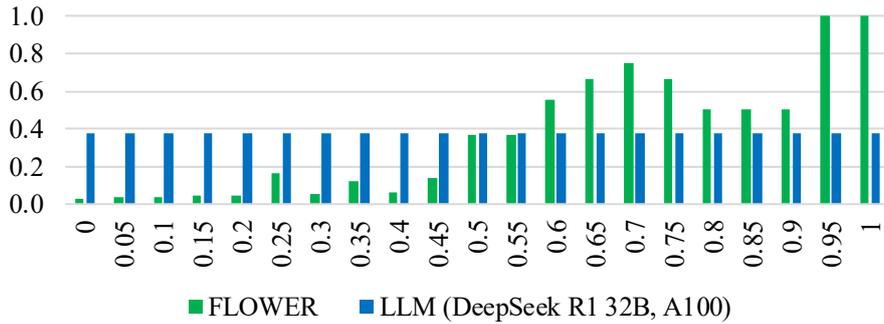

**Fig. 7.** Comparison of *FLOWER* and *baseline* in case of correct dependencies building.

The maximum improvement over the baseline was more than 2.6× when we used $confidence$ of 0.95 and 1 and calculated $accuracy$ metric (Eq. 6). We also measured that E2E latency for processing one table with FLOWER using dynamically sampled data is ~9.9 seconds while for LLM it took ~21.25 seconds that is ~2.15× improvements.

As for data storytelling FLOWER can brings advantages in terms of decreasing latency and context size using hierarchical dependencies for entities (Figure 8).

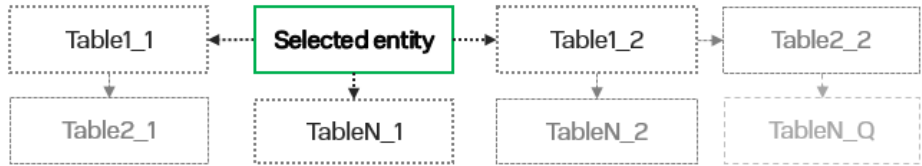

**Fig. 8.** Entities selection based on ER-model.

We compared LLM with ERD from our tool and vanilla LLM with all entities filled into prompt to understand how better it can describe columns in dataset (Table 3) with "Write few words about main *column* purpose in *table* using provided ER-model" as a common question template for each column and table.

Table 3. Data storytelling comparison (STATS).

| Technique | Context size | Latency |
| --- | --- | --- |
| LLM + FLOWER | *656* | *6.77* |
| LLM | 1220 | 8.03 |

Regarding answer quality of prompts, we measured it manually and observed latency and context improvements using ERD from FLOWER solution for entity description.

We noticed, that most of popular and SOTA benchmarks have only single schema, fixed naming and distribution pattern. Thus, it can't cover all use-cases for FLOWER, so own generating script BenchERD (Benchmark Entity Relationship Diagram) [16] was created and proposed for further testing on multiple schema workloads. It will allow us to name entities, choose distributions and flexibility to create DB for real-world scenarios.

## 5  Conclusion

In this paper we introduced FLOWER, an adaptive multidialectal tool that simplify and optimize routine tasks of constraint learning and data storytelling using own methods and decision-making system. Current solution is able to build explicit and implicit dependencies across all schemas in DBMS. FLOWER achieves 2.4× more accurate results of representing initial distribution using own dynamic sampling. It reaches up to 2.6× in constraint learning improvements with 2.15× E2E latency decrease. Accuracy improvements for data storytelling is 1.19× for latency and 1.86× for context decrease. Notable, that our solution supports 23 languages and can run smoothly on CPU, so GPU acceleration is possible, but not required. These results show that FLOWER is widely useful. Thus, it can be applicable for analysts, engineers, DBA and companies that want to create or improve ER-models, get some unseen insights from initial information using data storytelling or process with DB using natural language.